\documentclass[12pt]{elsarticle}

\makeatletter
\def\ps@pprintTitle{%
 \let\@oddhead\@empty
 \let\@evenhead\@empty
 \def\@oddfoot{\centerline{\thepage}}%
 \let\@evenfoot\@oddfoot}
\makeatother

\usepackage{amsmath,amssymb}
\usepackage{changepage}
\usepackage[dvipsnames]{xcolor}
\newcounter{bla}

\begin{document}
	
	\begin{frontmatter}
		
		
		
		\title{A Bayesian traction force microscopy method with automated denoising in a user-friendly software package}
		
		
		\author[a]{Yunfei Huang}
		\author[a]{Gerhard Gompper}
		\author[a]{Benedikt Sabass\corref{author}}
		
		\cortext[author] {Corresponding author.\\\textit{E-mail address:} b.sabass@fz-juelich.de}
		\address[a]{Theoretical Soft Matter and Biophysics, Institute of Complex Systems and Institute for Advanced Simulation, Forschungszentrum Juelich, 52425 Juelich, Germany}
		%\address[b]{Second Address}

		\begin{abstract}
			Adherent biological cells generate traction forces on a substrate that play a central role for migration, mechanosensing, differentiation, and collective behavior. The established
			method for quantifying this cell-substrate interaction is traction force microscopy (TFM). In spite of recent advancements, inference of the traction forces from measurements remains very sensitive to noise. However, suppression of the noise reduces the measurement accuracy and the spatial resolution, which makes it crucial to select an optimal level of noise reduction. Here, we present a fully automated method for noise reduction and robust, standardized traction-force reconstruction. The method, termed Bayesian Fourier transform traction cytometry, combines the robustness of Bayesian L2 regularization with the computation speed of Fourier transform traction cytometry. We validate the performance of the method with synthetic and real data. The method is made freely available as a software package with a graphical user-interface for intuitive usage.
		\end{abstract}
		
		\begin{keyword}
			Traction force microscopy, Bayesian inference.
		\end{keyword}
		
	\end{frontmatter}
	
	% \linenumbers

	%\section{}
	%\label{}
	
	\section{Introduction}
	Traction force microscopy (TFM) is a technique for measuring surface traction forces on an elastic substrate. Due to the unique possibilities offered by the technique, TFM enjoys wide popularity among biologists, materials scientists, and experimental physicists, see Refs.~\cite{polacheck2016measuring, schwarz2015traction, plotnikov2014high, roca2017quantifying, style2014traction, siedlik2016pushing,mulligan2018traction} for a non-comprehensive list of recent reviews. With TFM, one can record ``images'' and ``movies'' of the spatial distribution of traction forces on a surface. Moreover, TFM is essentially an imaging technique and does not require a perturbation of the sample. Therefore, the technique complements other techniques such as atomic force microscopy, optical tweezers, or the surface force apparatus. In materials science and physics, TFM has been used to measure interfacial forces during wetting, adhesion, rupture, and friction processes, see, e.g., Refs.~\cite{gerber2019wetting, xu2010imaging, maruthamutu}. However, the most important application of TFM is in biology, where it is being employed extensively for studying the mechanobiology of adherent cells. The traction patterns generated by adherent cells vary typically on a length scale of about one micrometer. The studied phenomena include cell migration and adhesion regulation~\cite{dembo1999stresses, butler2002traction,reilly2010intrinsic,betz2011growth,legant2013multidimensional,han2019formation}, three-dimensional collective cell organization~\cite{franck2011three,piotrowski2015three,aung20143d,steinwachs2016three},  and cell migration in wound-healing assays~\cite{trepat2009physical,brugues2014forces}, to name just a few. Overall, the length scales of traction patterns that have been studied with TFM range from below one micrometer in bacterial adhesion~\cite{sabass2017force, duvernoy2018asymmetric} to centimeters in propulsion waves of slugs and snails~\cite{lai2010mechanics}. 
	
\begin{figure}[!h]
		\centering
		\includegraphics[width=1\linewidth]{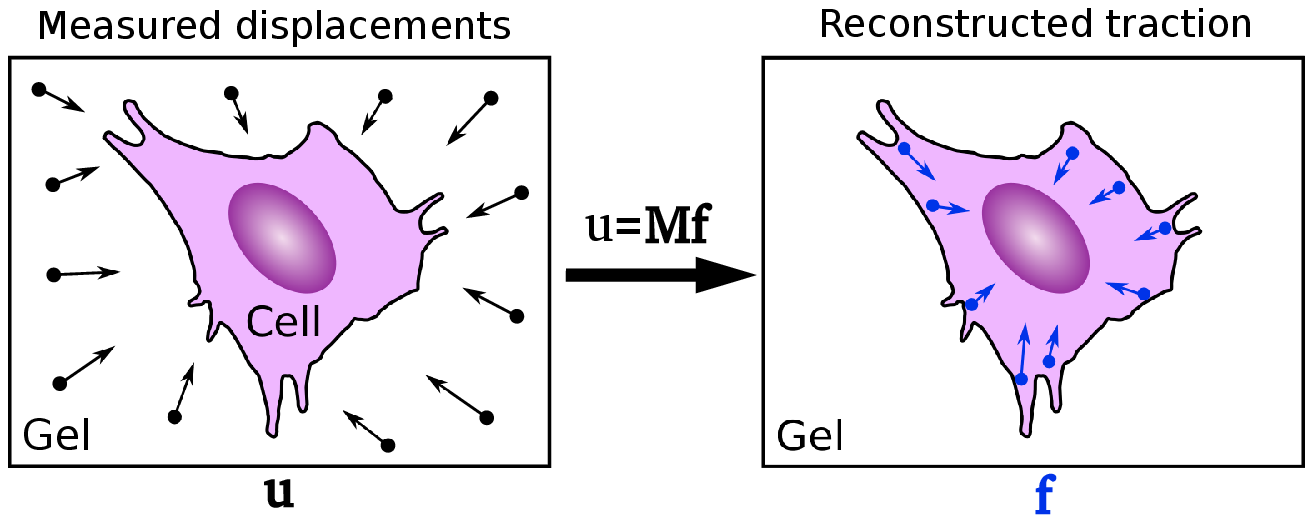}
		\caption{{\bf Schematic of traction force microscopy (TFM) to measure cellular traction on flat elastic substrates.}
			Adherent cells deform the substrate and the displacement field $\mathbf{u}$ is obtained by tracking markers within the gel. The traction force field $\mathbf{f}$ generated by the cell is calculated by inverting a linear equation system.}
		\label{fig1}
	\end{figure}

	A typical TFM setup for studying cell adhesion is sketched in Fig.~\ref{fig1}. Cells are placed on a flat elastic substrate which is usually synthetic, for example a soft polyacrylamide (PAA) gel, and contains fluorescent beads as fiducial markers.~\cite{sabass2008high} The live cells are imaged together with the fluorescent markers on a microscope. Surface traction results in local substrate deformations that can be monitored by tracking the motion of the beads in the substrate. The extraction of a discrete displacement field $\mathbf{U}$ describing the deformation of the substrates is usually done by comparing images of the fluorescent markers before and after removal of the adherent cells. Established methods for calculating the displacement fields include particle image velocimetry, single particle tracking, or optical flow tracking~\cite{plotnikov2014high,schwarz2015traction,holenstein2017high}.
Here and in the following, positions on the two-dimensional surface are described in a Cartesian system with coordinates $\mathbf{x}=(x_1,x_2)$. Using the planar deformations $(U_1,U_2)$ as input, a spatial map of traction $(F_1,F_2)$ on the gel surface can be mathematically reconstructed if out-of-plane forces are assumed to be negligible. For the reconstruction, the substrate is typically assumed to be a homogeneous, isotropic and linearly elastic half-space. Thus, the relation between the continuous displacement field $U_i(\mathbf{x})$ and the traction force field $F_j(\mathbf{x'})$ on the surface of substrates can be expressed as~\cite{landau1986theory}
	\begin{equation}
	U_i(\mathbf{x})=\int_{\Omega} \sum_{j=1}^{2} G_{ij}(\mathbf{x}-\mathbf{x'})F_j(\mathbf{x'})\,\mathrm{d}^2\mathbf{x'},
	\label{eq:1}
	\end{equation}
	where $G_{ij}(\mathbf{x})$ is the Green's function and $\Omega$ covers the whole substrate surface. The traction forces can be calculated in real space with finite element methods~\cite{yang2006determining,tang2014novel} or boundary element methods~\cite{sabass2008high,han2015traction}. To calculate the tractions based on Eq.~\eqref{eq:1} numerically, the integral equation needs to be discretized. In real space, one can employ linear shape functions~\cite{sabass2008high,han2015traction,huang2019traction} to write Eq.~\eqref{eq:1} as a linear matrix equation $\mathbf{u}=\mathbf{M}\mathbf{f}$, where the lower case letters $\mathbf{u}$ and $\mathbf{f}$ denote the discretized displacements and tractions. The coefficient matrix $\mathbf{M}$ results from integration of the shape functions. Such real-space methods are very flexible since they permit the study of various linear material responses encoded in the Green's function and spatial constraints are easily incorporated. However, accurate construction of the matrix $\mathbf{M}$ requires significant computation time on desktop machines. Alternatively, Eq.~\eqref{eq:1} can be solved in Fourier space by making use of the convolution theorem. This approach is called Fourier transform traction cytometry (FTTC)~\cite{butler2002traction,sabass2008high}. We employ a spatial wave vector $\mathbf{k}=(k_1,k_2)$ with absolute value $k=|\mathbf{k}|$. In standard FTTC, the integral Eq.~\eqref{eq:1} is written as $\tilde{u}_{i\mathbf{k}}=\{\sum_{j}\tilde{G}_{ij}\tilde{f}_{j}\}_\mathbf{k}$, where the tilde denotes the Fourier-transformed quantity. Using a matrix formulation analogous to the real-space expression, we have $\tilde{\mathbf{u}}=\tilde{\mathbf{M}}\tilde{\mathbf{f}}$ with $\tilde{\mathbf{M}}$ having a tri-diagonal structure. For conceptual clarity, in the following we will write the measurement noise in the recorded displacement explicitly as $\mathbf{s}$ in the real-space domain and as $\tilde{\mathbf{s}}$ in Fourier space. This noise can be estimated in the experiment by quantifying the variance of the measured displacements in absence of traction. The discretized equations then read 
	\begin{equation}
	\begin{cases}
	\mathbf{u}=\mathbf{M}\mathbf{f} + \mathbf{s} & \text{in real space},\\
	\tilde{\mathbf{u}}=\tilde{\mathbf{M}}\tilde{\mathbf{f}} + \tilde{\mathbf{s}} & \text{in Fourier space}.
	\end{cases}
	\label{eq:2}
	\end{equation}
	For traction force microscopy, either of these equations is employed to determine the tractions $\mathbf{f}$. The removal of noise is critical in most TFM methods. In real-space TFM calculations, the condition number of $\mathbf{M}$, defined as the ratio of the largest singular value to the smallest, is almost always much larger than unity, typically above $10^5$. $\mathbf{M}$ is therefore ill-conditioned which implies that small noise produces drastic changes in the calculated traction forces. For FTTC, spatially varying random noise occurs mainly at high spatial wave numbers. Hence, noise suppression can be achieved by suppressing high frequency data. In Ref.~\cite{huang2019traction}, we systematically tested a variety of traction reconstruction approaches based on Eq.~\eqref{eq:2}. The standard approach for solving the equation in real space is L2 regularization~\cite{dembo1996imaging, sabass2008high, houben2010estimating, schwarz2015traction}, which invokes a penalty on the traction magnitude to robustly suppress the effects of noise. With Fourier space methods, a low-pass filter is frequently employed to suppress noise in the displacement field before direct inversion of Eq.~\eqref{eq:2}~\cite{butler2002traction}. Alternatively, Fourier-space traction reconstruction can also be combined with L2 regularization, which conveys additional robustness~\cite{sabass2008high,plotnikov2014high,martiel2015measurement,holenstein2017high}.
	
	Virtually all standard methods for traction calculation require the implicit or explicit choice of a parameter that suppresses noise and leaves as much of the true signal conserved as possible. Within a Bayesian framework, this parameter choice can be rationalized by relating filter- or regularization parameters to prior distributions that represent prior knowledge about the data. Maximizing the likelihood of the prior distributions yields the corresponding optimal trade-off between noise suppression and faithful data reconstruction. Bayesian regularization has been used for example in astrophysics~\cite{suyu2006bayesian,ghosh2015bayesian} and mechanical structure monitoring~\cite{huang2014robust}. For inference of internal stress in a cell monolayer, an iterative maximum a posteriori estimation has been employed~\cite{Vincent}. TFM with Bayesian L2 regularization (BL2) was introduced in Ref.~\cite{huang2019traction} and is based on an established framework of Bayesian fitting~\cite{mackay1992bayesian}. Bayesian L2 regularization was first employed for real-space TFM methods since this variant allows comparison of a broad variety of approaches. For practical applications, however, calculations in Fourier domain have significant advantages in terms of robustness and speed. In this work, we present the corresponding method that we term Bayesian Fourier transform traction cytometry (BFTTC). We compare BFTTC with other methods such as classical L2 regularization, Bayesian L2 regularization in real-space (BL2), and regularized Fourier transform traction cytometry (FTTC). We find that BTTC is a computationally fast method that provides robust traction calculations without requiring manual adjustment of the noise-suppression level. We also provide a Matlab software package for BFTTC that is freely available for download. This software package is intended to provide a simple and robust solution for data analysis in the hands of experimentalists. A graphical user interface allows intuitive use of the program and little theoretical background knowledge is required. 
	
	\section{Methods and implementation}
	\subsection{Traction-displacement model}
	Assuming that the substrate is a semi-infinite half-space, the Green's function in Eq.~\eqref{eq:1} is given by the standard expression 
	\begin{equation}
	G_{ij}(\mathbf{x})=\frac{(1+\nu)}{\pi E}\left[(1-\nu)\frac{\delta_{ij}}{r}+\nu \frac{x_ix_j}{r^3}\right],
	\end{equation}
	where $E$ and $\nu$ represent the Young modulus and Poisson ratio, respectively. Here, $r=|\mathbf{x}|$ and $\delta_{ij}$ is the Kronecker delta function. Denoting the wave vector by $\mathbf{k}=(k_1,k_2)$ with absolute value $k=|\mathbf{k}|$, the Fourier-transformed Green function is given by 
	\begin{equation}	
	\tilde{G}_{ij\mathbf{k}}=\frac{2(1+\nu)}{E}\bigg[\frac{\delta_{ij}}{k}-\frac{\nu k_ik_j}{k^3}\bigg].
	\end{equation}
	The continuous traction and displacement fields are discretized by rectangular meshes where 
	$m$ and $n$ are the number of discretization nodes for tractions and displacements respectively. In the discretized equations~(\ref{eq:2}), the size of the displacement vector $\mathbf{u}$ is $2m\times1$ and size of the traction vector $\mathbf{f}$ is $2n\times1$ where the two vector components of the planar fields are concatenated. For the Fourier space methods, the displacement and traction fields are discretized with the same grid and we then have $m=n$.
	
	\subsection{Regularization}
	The classical approach to solve Eq.~\eqref{eq:2} is L2 regularization, which is also called Tikhonov regularization or ridge regression. L2 regularization is a robust procedure that suppresses noise and produces a smoothed traction field~\cite{huang2019traction}. Here, the residual $\|\mathbf{u}-\mathbf{Mf}\|^2_2 = (\mathbf{M}\mathbf{f}-\mathbf{u})^\mathrm{T}(\mathbf{M}\mathbf{f}-\mathbf{u})$ is minimized together with the solution norm $\lambda_2\|\mathbf{f}\|^2_2 = \lambda_2 \mathbf{f}^\mathrm{T} \mathbf{f}$. The factor $\lambda_2$ is called regularization parameter. The reconstructed traction $\hat{\mathbf{f}}$ satisfies
	\begin{equation}
	\hat{\mathbf{f}}=\underset{\mathbf{f}} {\text{argmin}}\big[\|\mathbf{M}\mathbf{f}-\mathbf{u}\|^2_2+\lambda_2\|\mathbf{f}\|^2_2\big].
	\label{eq:3}
	\end{equation}
	This approach can be employed for real-space TFM and in Fourier space, where the square norms can be calculated conveniently with Parseval's theorem.
	The proper choice of the regularization parameter $\lambda_2$ is critical in case accurate traction calculations are required. A popular heuristic for choosing the regularization parameter is based on a double-logarithmic plot of the solution residual \textit{vs.} the traction norm for varying $\lambda_2$. Often, the plotted line resembles an ``L'' shape and the regularization parameter is chosen to lie in the corner of this curve, thus providing a trade-off between faithful reconstruction and smoothing~\cite{hansen2007regularization}. However, this ``L-curve criterion'' is often of little use, in particular when the corner is either absent or cannot be localized precisely on the double-logarithmic scale. Moreover, it has also been shown that the L-curve criterion can fail systematically~\cite{hanke1996limitations,Vogel1996}. Therefore, the L-curve criterion is often  complemented with other methods for finding the regularization parameter, such as cross-validation~\cite{huang2019traction}. In any case, a manual parameter variation is mandatory to check the validity of the solution.
	
	\subsection{Bayesian Fourier transform traction cytometry}
	Bayesian methods can be used to regularize data in a systematic and automated way. Our approach is based on an established iterative inference procedure~\cite{mackay1992bayesian}. In the first step, a model is fitted to the data. In the second step, the evidence for the chosen model is calculated. Traction computations with Bayesian L2 regularization (BL2) were first introduced as a real-space approach in Ref.~\cite{huang2019traction}. Here we describe the adaptation of this method to Fourier-space traction calculation. It is assumed that the noise $\mathbf{s}$ in Eq.~\eqref{eq:2} has a Gaussian distribution with vanishing mean and a variance of $1/\beta$ . Therefore, given a traction vector $\mathbf{f}$, the likelihood of measuring a particular $2m\times1$ displacement vector $\mathbf{u}$ is  
	\begin{equation}
	p(\mathbf{u}|\mathbf{f},\beta)=\frac{\exp[-\beta E_\text{u}]}{Z_\text{u}}=\frac{\exp[-\beta(\mathbf{M}\mathbf{f}-\mathbf{u})^\mathbf{T}(\mathbf{M}\mathbf{f}-\mathbf{u})/2]}{Z_\text{u}},
	\label{eq:4}
	\end{equation}
	where $Z_\text{u}=(2\pi/\beta)^m$. As a prior distribution for the $2n\times1$ vector of traction forces $\mathbf{f}$ we choose a Gaussian distribution with variance $1/\alpha$ as
	\begin{equation}
	p(\mathbf{f}|\alpha)=\frac{\exp[-\alpha E_\text{f}]}{Z_\text{f}}=\frac{\exp[-\alpha\mathbf{f}^\mathbf{T}\mathbf{f}/2]}{Z_\text{f}},
	\label{eq:5}
	\end{equation}
	where $Z_\text{f}=(2\pi/\alpha)^n$. According to Bayes' rule, the posterior distribution of $\mathbf{f}$ is given by 
	\begin{equation}
	p(\mathbf{f}|\mathbf{u},\alpha,\beta)=\frac{p(\mathbf{u}|\mathbf{f},\beta)p(\mathbf{f}|\alpha)}{p(\mathbf{u}|\alpha,\beta)}=\frac{\exp[-K(\mathbf{f})]}{p(\mathbf{u}|\alpha,\beta)Z_\text{u}Z_\text{f}},
	\label{eq:6}
	\end{equation}
	where $K(\mathbf{f})=\beta E_\text{u}+\alpha E_\text{f}$ and $p(\mathbf{u}|\alpha,\beta)=\int \mathrm{d}^{2n} \mathbf{f} \exp[-K(\mathbf{f})]/(Z_\text{u}Z_\text{f})$. To find the traction vector with the highest posterior probability, we maximize $p(\mathbf{f}|\mathbf{u},\alpha,\beta)$ with respect to $\mathbf{f}$. The calculation yields $\mathbf{f}_{\text{MP}}= \underset{\mathbf{f}} {\text{argmin}}\big[\beta\|\mathbf{M}\mathbf{f}-\mathbf{u}\|^2_2/2+\alpha\|\mathbf{f}\|^2_2/2\big]$, which is equivalent to our formula for L2 regularization, Eq.~\ref{eq:3}, when $\lambda_2=\alpha/\beta$ ~\cite{plotnikov2014high}. 
	
	Next, the values of the hyperparameters $\alpha$ and $\beta$ have to be determined. In principle, both values can be found by maximizing the evidence $p(\alpha,\beta|\mathbf{u})$ that depends on the measured displacements $\mathbf{u}$. However, the noise variance $1/\beta$ can also be estimated directly from the measurement uncertainty. Thereby, the maximization of $p(\alpha,\beta|\mathbf{u})$ can be reduced to a robust one-dimensional search for the optimal value of $\alpha$. Bayes' law yields $p(\alpha,\beta|\mathbf{u})=p(\mathbf{u}|\alpha,\beta)p(\alpha,\beta)/p(\mathbf{u})$. We next assume a uniform prior $p(\alpha,\beta)\simeq \mathrm{const.}$ and note that $p(\mathbf{u})$ does not play a role for the optimization. Thus, we only need to maximize $p(\mathbf{u}|\alpha,\beta)\sim\int \mathrm{d}^{2n} \mathbf{f} \exp[-K(\mathbf{f})]$ with respect to $\alpha$. The integral can be analytically calculated by completing the square. 
	On defining  $\mathbf{A}=\alpha\mathbf{I}+\beta\mathbf{M}^{\mathbf{T}}\mathbf{M}$ one finds
	\begin{equation}
	p(\mathbf{u}|\alpha,\beta)=\frac{\int \mathrm{d}^{2n} \mathbf{f} \exp[-K(\mathbf{f})]}{Z_\text{u}Z_\text{f}}=\frac{(2\pi)^n(\det\mathbf{A})^{-1/2}}{Z_{\text{u}}Z_{\text{f}}}\exp[-\mathbf{K}(\mathbf{f}_\text{MP})]. 
	\label{eq:7}
	\end{equation}
	Since $\mathbf{f}_\text{MP}$ and $\mathbf{A}$ both depend on $\alpha$, the maximization of Eq.~\eqref{eq:7} with respect to $\alpha$ needs to be done iteratively. This iteration can be sped up by performing the calculations in Fourier space. For notational clarity, we will write Fourier-space variables and derived quantities with a tilde. The Fourier-transformation of $\mathbf{f}_{\text{MP}}$ yields $\mathbf{\tilde{f}}_{\text{MP}} = (\mathbf{\tilde{M}^{\dagger}}\mathbf{\tilde{M}}+\alpha/\beta\mathbf{I})^{-1}\mathbf{\tilde{M}^{\dagger}}\mathbf{\tilde{u}}$~\cite{sabass2008high}, where the complex transpose is indicated by a $\dagger$. Parseval's theorem allows convenient expression of Eq.~\eqref{eq:7} through Fourier-space variables. We have $\tilde{E}_\text{u}=(\tilde{\mathbf{M}}\tilde{\mathbf{f}}-\tilde{\mathbf{u}})^\mathbf{\dagger}(\tilde{\mathbf{M}}\tilde{\mathbf{f}}-\tilde{\mathbf{u}})/(2m)$, $\tilde{E}_\text{f}=\tilde{\mathbf{f}}^\mathbf{\dagger}\tilde{\mathbf{f}}/(2 n)$, and $\tilde{\mathbf{A}}=\alpha\mathbf{I}/n+\beta\tilde{\mathbf{M}}^{\mathbf{\dagger}}\tilde{\mathbf{M}}/m$. Using these expressions, the logarithm of the evidence, cf. Eq.~\eqref{eq:7}, can be written as
	\begin{equation}
	\log p(\mathbf{\tilde{u}}|\alpha,\beta)=-\beta\tilde{E}_\text{u}(\mathbf{\tilde{f}}_\text{MP})-\alpha\tilde{E}_\text{f}(\mathbf{\tilde{f}}_\text{MP})-\frac{1}{2}\log(\det\tilde{\mathbf{A}})+n\log\alpha+m\log\beta-m\log(2\pi).
	\label{eq:8}
	\end{equation}
	This expression is evaluated numerically. The calculation of $\log(\det\tilde{\mathbf{A}})$ is done by a Cholesky decomposition of the positive matrix $\tilde{\mathbf{A}}=\mathbf{L}\mathbf{L}^T$ as $\log(\det(\mathbf{LL^T}))=2\log\Pi_iL_{ii}=2\Sigma_i\log(L_{ii})$~\cite{huang2019traction}. To determine the value of $\alpha = \hat{\alpha}$ that maximizes $\log p(\mathbf{\tilde{u}}|\alpha,\beta)$ we employ a golden-section search. Finally, the L2 regularization parameter follows as $\hat{\lambda}_2=\hat{\alpha}/\beta$.

The calculation of the parameter value $\hat{\lambda}_2$ requires a well-defined maximum of the logarithmic evidence as a function of $\alpha$. To assess whether this maximum exists, we investigate the condition $\frac{\mathrm{d}}{\mathrm{d}\alpha}\log p(\mathbf{\tilde{u}}|\alpha,\beta)=0$. For evaluation of the derivatives of $\tilde{E}_\text{u}(\mathbf{\tilde{f}}_\text{MP})$ and $\tilde{E}_\text{f}(\mathbf{\tilde{f}}_\text{MP})$ we use that $n = m$ and that $\mathbf{\tilde{M}}$ commutes with $(\mathbf{\tilde{M}^{\dagger}}\mathbf{\tilde{M}}+\alpha/\beta\mathbf{I})^{-1}$ since the Fourier-transformed Green's function is a real, symmetric matrix. A straightforward calculation yields $\frac{\mathrm{d}}{\mathrm{d}\alpha}\tilde{E}_\text{u}(\mathbf{\tilde{f}}_\text{MP}) = -\lambda \frac{\mathrm{d}}{\mathrm{d}\alpha}\tilde{E}_\text{f}(\mathbf{\tilde{f}}_\text{MP})$. Therefore, the condition determining the maximum becomes $0=\frac{\mathrm{d}}{\mathrm{d}\alpha}\log p(\mathbf{\tilde{u}}|\alpha,\beta) = -\tilde{E}_\text{f}(\mathbf{\tilde{f}}_\text{MP}) -\frac{1}{2 n}\mathrm{Tr}[\tilde{\mathbf{A}}^{-1}] +\frac{n}{\alpha}$. We next perform a symbolic eigenvalue decomposition of $\mathbf{\tilde{M}}$ and denote the eigenvalues by $\{m_i\}$, the matrix of eigenvectors by $\mathbf{V}^{T}$, and define $\hat{u}_i = V_{ij}\tilde{u}_j$. The condition determining the maximum of the logarithmic evidence then reads
\begin{equation}
\frac{1}{2 n}\sum_{i=1}^{2 n} \frac{\beta\hat{u}_i^{\dagger}\hat{u}_i\, m_i^2}{(m_i^2 + \lambda_2)^2}
=\frac{1}{2}\sum_{i=1}^{2 n}\frac{m_i^2}{\lambda_2(m_i^2 + \lambda_2)}.
\label{eq_determine_max}
\end{equation}
Solutions exist if the functions of $\lambda_2$ on the left hand side and on the right hand side of Eq.~(\ref{eq_determine_max}) cross each other. Both functions decrease monotonously with $\lambda_2$. However, for $\lambda_2 \rightarrow 0$ the left hand side remains finite while the right hand side diverges. Thus, Eq.~(\ref{eq_determine_max}) has a real solution if the left hand side becomes bigger than the right hand side for any $\lambda_2\geq0$. In the limit of $\lambda_2 \rightarrow \infty$, the condition for the occurence of a maxium becomes $\frac{1}{n}\sum_{i=1}^{2 n} \beta\hat{u}_i^{\dagger}\hat{u}_i\, m_i^2/\left(\sum_{j=1}^{2 n}m_j^2\right) \geq 1$. For the TFM data, we find that the values of $\hat{u}_i^{\dagger}\hat{u}_i$ roughly decrease with decreasing squared eigenvalues $m_i^2$ since the displacement magnitudes typically decrease with higher Fourier modes, as do the entries of $\tilde{\mathbf{M}}$. Assuming that the approximate ordering of $m_i^2$ and $\hat{u}_i^{\dagger}\hat{u}_i$ holds strictly, we can invoke Chebyshev's sum inequality to obtain $\frac{1}{n}\sum_{i=1}^{2 n} \beta\hat{u}_i^{\dagger}\hat{u}_i\, m_i^2/\left(\sum_{j=1}^{2 n}m_j^2\right) \geq \frac{1}{n^2}\sum_{i=1}^{n} \beta\hat{u}_i^{\dagger}\hat{u}_i$. Since for all reasonable TFM datasets the mean squared displacement is larger than the noise variance, we expect that $\frac{1}{n^2}\sum_{i=1}^{n} \beta\hat{u}_i^{\dagger}\hat{u}_i = \frac{2 \beta}{n} \sum_{i=1}^{n} u_i^2 >1$. Therefore, the condition for the occurrence of a maximum in $\log p(\tilde{\mathbf{u}}|\alpha,\beta)$ should be fulfilled for some $\lambda_2 >0$. The resulting maximum is unique due to defined signs of the $\lambda_2$ derivatives of Eq.~(\ref{eq_determine_max}).
In summary, a semi-quantitative argument supports the existence of a unique maximum of the logarithmic evidence $\log p(\tilde{\mathbf{u}}|\alpha,\beta)$ when appropriate TFM data is used. In our tests, a maximum was found for all datasets.

	\subsection{Generation of synthetic test data}
	To confirm that the Bayesian approach yields a correct estimate for the regularization parameter we employ synthetic data sets with known properties. In our first test series, we generate random traction fields by drawing individual traction vectors from Gaussian distributions with fixed variances, as illustrated in Fig.~\ref{fig3} (a-i) and (a-ii). The traction field is produced on a $50\times50$ grid with a Young modulus of $E=10\,\rm{kPa}$ and a Poisson ratio of $\nu=0.3$. For example, we employ a Gaussian traction distribution with a variance of $10^{4}\, \mathrm{Pa}^2$ and therefore $\alpha=10^{-4}\,\mathrm{Pa}^{-2}$. After calculation of the displacements from the traction, Gaussian noise with a variance of $10^{-4}\, \mathrm{Pix}^2$ is added, thus $\beta=10^{4}\, \mathrm{Pix}^{-2}$.
	In the second test series, we construct synthetic data to study the reconstruction quality for localized traction patterns. As in previous work~\cite{sabass2008high,huang2019traction}, {we assume that the traction is localized in circular spots, each having a constant traction magnitude. For every individual spot, the step-like traction profile can be integrated analytically to produce a displacement field. Due to the linearity of the problem, displacements from different spots can be added to produce the final result. Explicit formulas for the displacement field are provided in the supplementary of Ref.~\cite{huang2019traction}.}
For generation of this data, we fix the Young modulus $E=10\,\rm{kPa}$ and the Poisson ratio $\nu=0.3$. The traction patterns consist of $10-20$ circular traction spots, as illustrated in Fig.~\ref{fig4}(a). The diameter of the spots is $2\,\mathrm{\mu m}$ and the mesh size of the reconstructed traction is $0.5\,\mathrm{\mu m}$. The traction force magnitude in the spots is  randomly chosen in the range $[0-700]\,\mathrm{Pa}$ and the sum of the x- and y components of the traction forces vanishes. To simulate the measurement uncertainty, Gaussian noise is added after calculation of the displacement field. The noise variance in the different samples is between 2\% and 8\% of the maximum absolute displacement value.
	
	\subsection{Reconstruction quality measures}
	For the synthetic test data with circular spots the traction force is exactly known. Therefore, we can qualitatively calculate the reconstruction errors. Here, we use four different error measures introduced in our previous work{~\cite{sabass2008high,huang2019traction}}. To provide simple definitions of the error measures, we rewrite the $2m \times 1$ traction vector $\mathbf{f}$ as a $m \times 2$ traction vector with the values $\mathbf{t} = \{t_x,t_y\}$ at every grid node. Real traction and reconstructed traction are denoted by $\mathbf{t}^{\text{true}}$ and $\mathbf{t}^{\text{recon}}$, respectively.
	\begin{itemize}
		\item The Deviation of Traction Magnitude at Adhesions (DTMA) is defined as 
		\begin{equation}
		\text{DTMA}=\frac{1}{N_i}\sum_{i}\frac{\text{mean}_j\left(\|\mathbf{t}^{\text{recon}}_{j,i}\|_2-\|\mathbf{t}^{\text{true}}_{j,i}\|_2\right)}{\text{mean}_j\left(\|\mathbf{t}^{\text{true}}_{j,i}\|_2\right)},
		\label{DTMA}
		\end{equation}
		where $N_i$ is the number of circular traction patches and the index $i$ runs over all patches. The index $j$ runs over all traction vectors in one patch. The DTMA lies between $-1$ and $1$ where 0 indicates a perfect average traction recovery and a negative or positive value implies underestimation or overestimation, respectively.
		
		\item The Deviation of Traction Magnitude in the Background (DTMB) is the normalized difference between the reconstructed and real traction magnitude outside the circular patches
		\begin{equation}
		\text{DTMB}=\frac{\text{mean}_k\left(\|\mathbf{t}^{\text{recon}}_{k}\|_2-\|\mathbf{t}^{\text{true}}_{k}\|_2\right)}{\frac{1}{N_i}\sum_{i}\text{mean}_j\left(\|\mathbf{t}^{\text{true}}_{j,i}\|_2\right)},
		\label{DTMB}
		\end{equation}
		where the index $k$ runs over all traction vectors outside the patches. The DTMB lies in the range $[0,1]$ and a value close to $0$ indicates low background noise in the reconstructed traction.
		
		\item The Signal to Noise Ratio (SNR) is defined in this context as
		\begin{equation}
		\text{SNR}=\frac{\frac{1}{N_i}\sum_i \text{mean}_j(\|\mathbf{t}^{\text{recon}}_{j,i}\|_2)}{ \text{std}_k(\mathbf{t}^{\text{recon}}_{k})}.
		\label{SNR}
		\end{equation}
		The index $k$ runs over all traction vectors outside the patches while $j$ is the index of each traction vector in the patch $i$. The SNR measures the detectability of a real signal within a noisy background. Its value ranges from 0 to infinity where a SNR that is much larger than unity indicates a good separation between traction and noise. 
		
		\item The Deviation of the traction Maximum at Adhesions (DMA) measures how peak values of the traction over- or underestimate the true value. The quantity is defined as
		\begin{equation}
		\text{DMA}=\frac{1}{N_A}\sum_i\frac{ \left[\text{max}_j(\|\mathbf{t}^{\text{recon}}_{j,i}\|_2)-\text{max}_j(\|\mathbf{t}^{\text{true}}_{j,i}\|_2)\right]}{\text{max}_j(\|\mathbf{t}^{\text{true}}_{j,i}\|_2)},
		\label{DMA}
		\end{equation} 
		where the maximum traction vector with index $j$ is calculated for each traction patch separately. A DMA of 0 indicates that the local traction maxima in the reconstruction and in the original data are equal. Positive or negative values of the DMA imply that the maximum of traction is overestimated or underestimated. 
	\end{itemize}
	\subsection{Software for traction force calculation}
	We provide a Matlab software package containing the presented Fourier-space methods for calculating traction forces. {Note that the program requires
the input of substrate-deformation data. Usually, substrate deformations are quantified by measuring the lateral displacements of fluorescent marker beads in a stressed substrate with respect to the marker positions recorded in a stress-free state. The standard computational image analysis method for this task is called particle image velocimetry (PIV) and various well-established software packages are available~\cite{PIV-Lab,PIV-mpiv,PIV-vennemann}. Once the displacement data has been extracted,} our program can be used to calculate the traction forces with standard L2 regularization or with Bayesian L2 regularization in Fourier space. The software is split into a routine for loading data and two routines for TFM. The routine ``get input data" allows the user to select folders containing the data for the measured displacements, the noise, and for images. The required data structure in the file with the displacement data is illustrated in Fig.~\ref{fig2} (a).
	Parameters of the experimental setup, including the Young modulus and the Poisson ratio, also need to be provided. Next, the user can choose between ``Regularization" and ``Bayesian regularization", as shown in Fig.~\ref{fig2} (b) and (c). Selecting ``Regularization" allows the choice of a regularization parameter, which is then held fixed for the whole sequence of images that are analyzed in the data set. For ``Bayesian regularization", an optimal regularization parameter is selected automatically from the data set and the noise variance. The standard deviation of the noise can either be provided as an input or can be determined by manually selecting an image region that is far away from the cell, as illustrated in Fig.~\ref{fig2} (c). Once selected, the region used for determining the noise remains the same throughout the whole data set of multiple images. After pressing ``Analyze sequence'' the results are calculated and saved in automatically named files, see  Fig.~\ref{fig2}~(c). \\
Since the regularization parameter $\lambda_2$ depends in our framework on the noise and the traction magnitudes, it should be adapted if the signal-to-noise level changes significantly. However, note that a change of the parameter within one image sequence is not always necessary, which reduces the computational effort and may be advantageous for data postprocessing.

	\section{Results}
	\subsection{Validation of the method with synthetic data}
\begin{figure}[!h]
		\centering
		\includegraphics[width=1\linewidth]{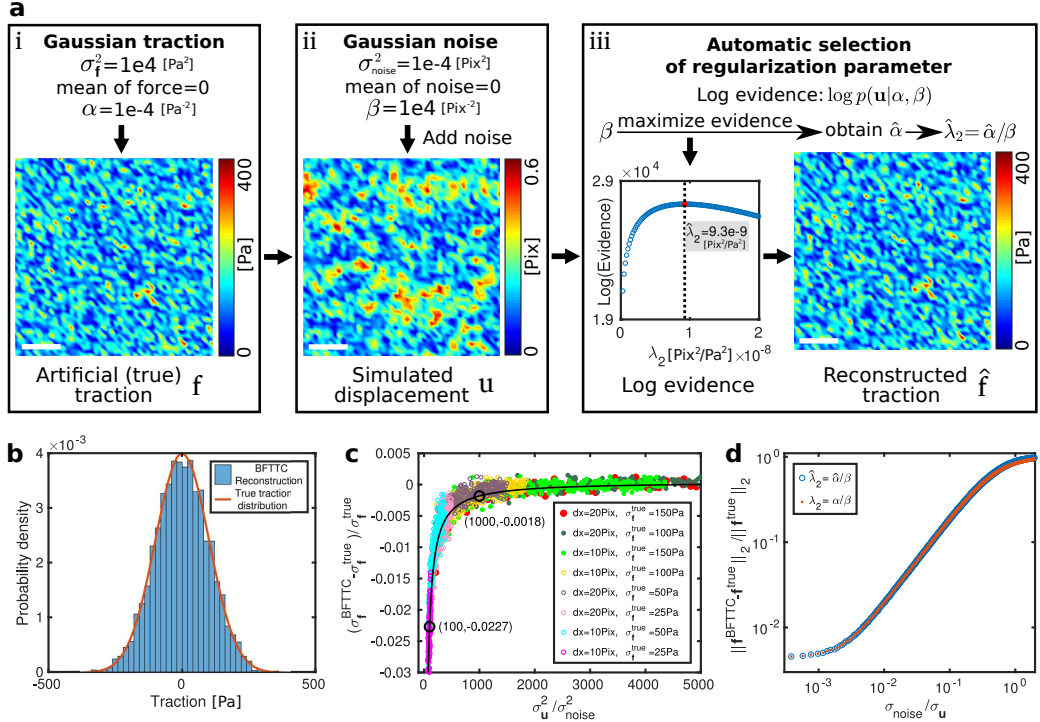}
		\caption{{\bf Validation of the Bayesian method for regularization parameter choice.} (a,i)~Traction force vectors, discretized on a quadratic mesh, are randomly chosen from a Gaussian distribution with fixed variance $\sigma^2_{\mathbf{f}}$. Space bar:~$100\,\mathrm{Pix}=10$~grid spacings on a $50 \times 50$ mesh. (a,ii)~Using the prescribed traction as input, a displacement field is calculated and Gaussian noise with a variance $\sigma^2_{\mathrm{noise}}$ is added. (a-iii)~The regularization parameter is determined by localizing the maximum in the log evidence curve and traction forces are subsequently calculated. (b)~Histogram of the tractions for the sample shown in (a). In the limit of weak noise, the histogram of the reconstructed traction matches the true traction distribution. (c)~Relative difference between the standard deviation of the measured traction distribution $\sigma_{\mathbf{f}}^{\text{BFTTC}}$ and the width of true traction distribution $\sigma_{\mathbf{f}}^{\text{true}}$. The grid mesh sizes are denoted by $\mathrm{dx}$. $\sigma^2_{\mathrm{\mathbf{u}}}$ is the variance of the synthetic displacement data prior to corruption with noise. Increasing the noise level produces a measured (posterior) traction distribution that no longer agrees with the true traction distribution. (d)~Mean error of the reconstructed traction as a function of the relative measurement uncertainty $\sigma_{\text{noise}}/\sigma_{\text{u}}$. The Bayesian estimate for the regularization parameter $\hat{\lambda}_2$ and the optimal regularization parameter $\alpha/\beta$ produce comparable errors for all noise levels.}
		\label{fig3}
	\end{figure}

	To check whether the proposed method actually finds the correct regularization parameter, synthetic data sets with exactly known underlying distributions are required. Therefore, we create random traction patters with traction vectors at each grid point drawn from a Gaussian distribution. Exemplary data is shown in Fig.~\ref{fig3}~(a-i). The calculated displacement field is then corrupted with a controlled level of noise, see Fig.~\ref{fig3}(a-ii). For the reconstruction, we search for the hyperparameter $\alpha$ that maximizes the log-likelihood function, Eq.~(\ref{eq:8}). As illustrated in Fig.~\ref{fig3}~(a-iii), $\log p(\mathbf{u}|\alpha,\beta)$ has a unique, clear maximum. The regularization parameter determined from the optimization compares favorably with the true optimal parameter resulting from the distributions used for simulating the data, here   $\hat{\lambda}_2=9.3\times 10^{-9}\,\mathrm{Pix}^2/\mathrm{Pa}^2 \simeq \alpha/\beta=10^{-8}\,\mathrm{Pix}^2/\mathrm{Pa}^2$. Visual comparison of the traction patterns in Figs~\ref{fig3}~(a-i,a-iv), as well as a comparison of the traction distributions in Fig.~\ref{fig3}~(b), confirm that the Bayesian traction reconstruction yields correct results. Note that the measured (posterior) traction distribution does not agree with the original traction distribution when the noise magnitude is large. This  fact is a result of the deviation of the posterior probability distribution, Eq.~(\ref{eq:6}), from the prior probability distribution. In Fig.~\ref{fig3}~(c), we illustrate the difference between the measured traction distribution and the original traction distribution for the synthetic data. The relative difference of the traction standard deviations is plotted against the variance of the noise-free displacement field divided by the noise variance, $\sigma^2_{\text{u}}/\sigma^2_{\text{noise}}$. The relative difference of the standard deviation of the measured posterior and the original traction distribution scales with the relative noise variance. Figure~\ref{fig3}~(d) illustrates how the measurement uncertainty affects the mean traction error. For the experimentally relevant regime of measurement uncertainties, $0.01 \gtrsim  \sigma_{\text{noise}}/\sigma_{\text{u}} \gtrsim 0.1$, the relative mean traction error is almost proportional to the relative measurement uncertainty $\sigma_{\text{noise}}/\sigma_{\text{u}}$. For very low measurement noise, the mean traction error is dominated by numerical inaccuracy and aliasing effects. Note that the Bayesian estimate for the regularization parameters $\hat{\lambda}_2$ produces errors that are close to the optimal errors resulting from regularization with the known parameters $\alpha/\beta$ for synthetic data. 
	
	\subsection{Quality assessment of traction reconstruction with BFTTC}
\begin{figure}[!h]
		\centering
		\includegraphics[width=1\linewidth]{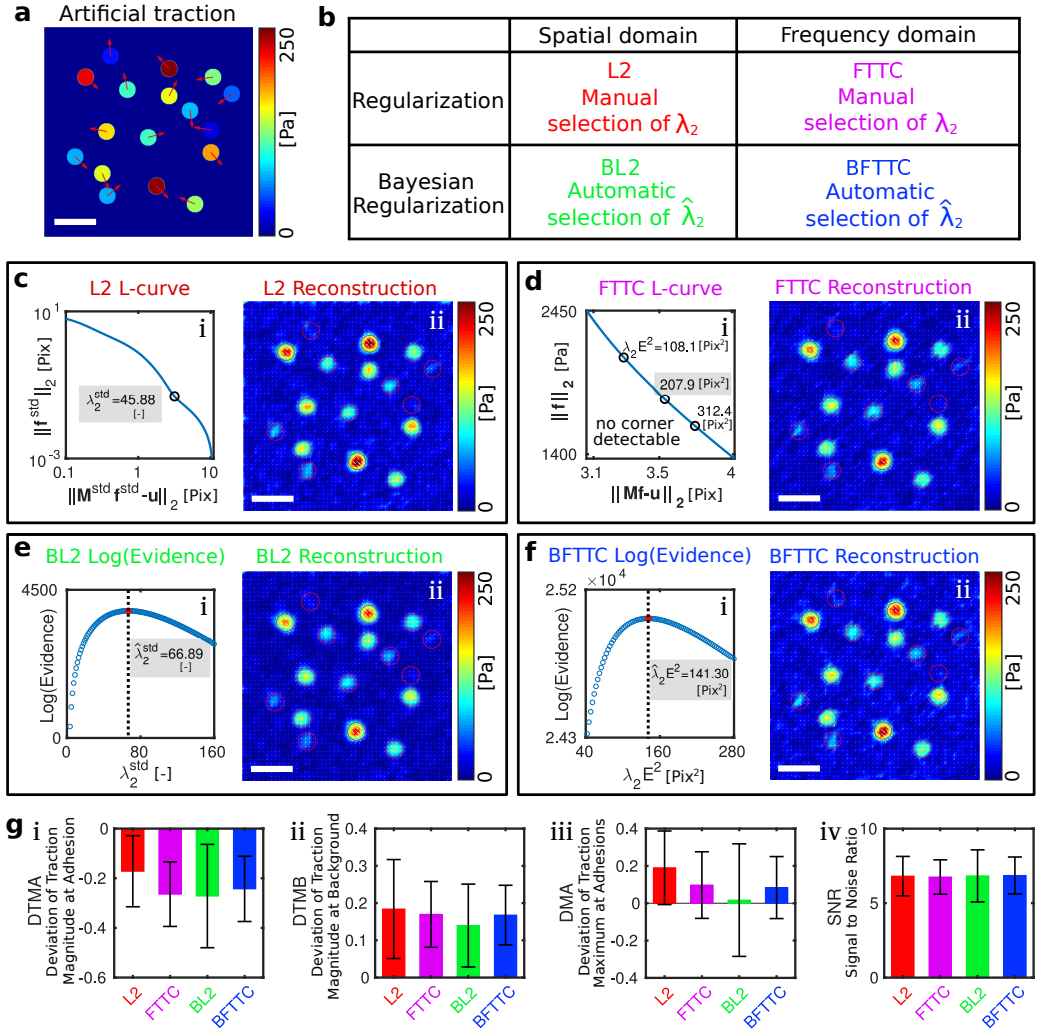}
		\caption{{\bf Reconstruction quality of BFTTC compared to other regularization methods.}
			(a)~Synthetic traction force pattern that is used for testing the reconstruction. Space bar:~$5\,\mu\rm{m}$. (b)~Tabulated overview of the compared traction reconstruction methods. (c)~Classical traction reconstruction in real space with L2 regularization. The L-curve shows a slight ``corner'', which is used to determine the value of the regularization parameter. Note that calculations in real space are done with standardized data~\cite{huang2019traction}, which renders the regularization parameter dimensionless. (d)~Bayesian L2 regularization (BL2) in real space determines the regularization parameter value automatically. The automatically determined regularization parameter is close to the one predicted in (c) from the L-curve. (e)~Classical, regularized Fourier transform traction cytometry (FTTC). The L-curve does not show a ``corner'', which makes it difficult to determine an appropriate regularization parameter. (f) Bayesian Fourier transform traction cytometry (BFTTC) determines an optimal regularization parameter automatically. (f)~Comparison of the reconstruction quality measures in $8$ synthetic data sets; error bars are the standard deviations of the samples. The reconstruction accuracy of all four methods is found to be similar.}
		\label{fig4}
	\end{figure}

	To quantify the reconstruction quality for localized traction patterns, {we construct} synthetic data consisting of circular spots of constant traction as shown in Fig.~\ref{fig4}~(a). We employ two classical methods where the regularization parameter value is selected by the L-curve criterion, namely a real space calculation with L2 regularization and regularized Fourier transform traction cytometry (FTTC). The results are compared with the corresponding parameter-free approaches, namely Bayesian L2 regularization (BL2) in the real-space domain and Bayesian Fourier transform traction cytometry (BFTTC), see Fig.~\ref{fig4}~(b). For the real-space TFM results shown exemplarily in Fig.~\ref{fig4}~(c), the L-curve can have a visible corner.  Note that the calculations in real space are done with standardized data~\cite{huang2019traction}, which renders the regularization parameter dimensionless. For the Bayesian real-space approach, illustrated in Fig.~\ref{fig4}~(e), the logarithmic evidence always exhibits a clear maximum in our experience. The resulting optimal regularization parameter is usually close to the value from the L-curve criterion. However, in the Fourier-space approach, illustrated with the example in Fig.~\ref{fig4}~(d), the L-curve often does not show a clear corner and it becomes challenging to select an appropriate regularization parameter. This weakness of the Fourier-space approach is overcome with BFTTC. As illustrated in Fig.~\ref{fig4}~(f), the logarithmic evidence calculated in BFTTC has a pronounced maximum, which provides a clear criterion for the automated choice of the optimal regularization parameter. To generate statistics on the performance of the different methods, we next record the traction reconstruction quality in $8$ separate tests with different traction magnitudes and patterns. The resulting error norms show that all four methods offer similar traction reconstruction accuracies, see Figs.~\ref{fig4}(e)(i-iv). The most noticeable reconstruction errors are an underestimation of mean traction (negative DTMA) and a pronounced traction background (positive DTMB)~\cite{huang2019traction}. The similarity in reconstruction accuracy is expected because all methods are based on L2 regularization and also make use of the same spatial grid for discretization. However, the numerical effort required for the four methods is very different. Table~\ref{table1} summarizes the computation time required for building the coefficient matrices $\mathbf{M}$ or $\tilde{\mathbf{M}}$ and for reconstructing the traction forces. While $\tilde{\mathbf{M}}$ is rapidly built in Fourier space, the assembly of a large coefficient matrix $\mathbf{M}$ in real space requires can require many hours. Inferring the optimal regularization parameter requires additional computation time. Overall, real-space methods are not prohibitively slow but quite impractical for every-day use by experimental scientists. BFTTC, however, requires acceptable computation times ranging from seconds to a few minutes.
	\begin{table}[!ht]
		%	\begin{adjustwidth}{-2.25in}{0in}
		\centering
		\caption{
			{\bf Computation time for different methods}}
		\begin{tabular}{|c|c|c|c|c|}	
			\hline
			Reconstruction method & L2 & BL2 & FTTC & BFTTC \\ 
			\hline
			\rule{0pt}{10pt} Building of $\mathbf{M}$ or $\tilde{\mathbf{M}}$ & \multicolumn{2}{c|}{23.3 h} & \multicolumn{2}{c|}{ 0.07 s }\\
			\hline
			Traction reconstruction & 67.4 s & 338.8 s & 0.06 s  &  3.1 s  \\ 
			\hline
		\end{tabular}
		\begin{flushleft} The employed data set consists of a rectangular grid with 2500 displacement and traction vectors. Benchmark tests
			were done on a desktop computer equipped with 16 GB RAM and an Intel I5-7500 CPU (3.40 GHz).
		\end{flushleft}
		\label{table1}
		%	\end{adjustwidth}
	\end{table}
	
	\subsection{Application of BFTTC to experimental data}
\begin{figure}[!h]
		\centering
		\includegraphics[width=1\linewidth]{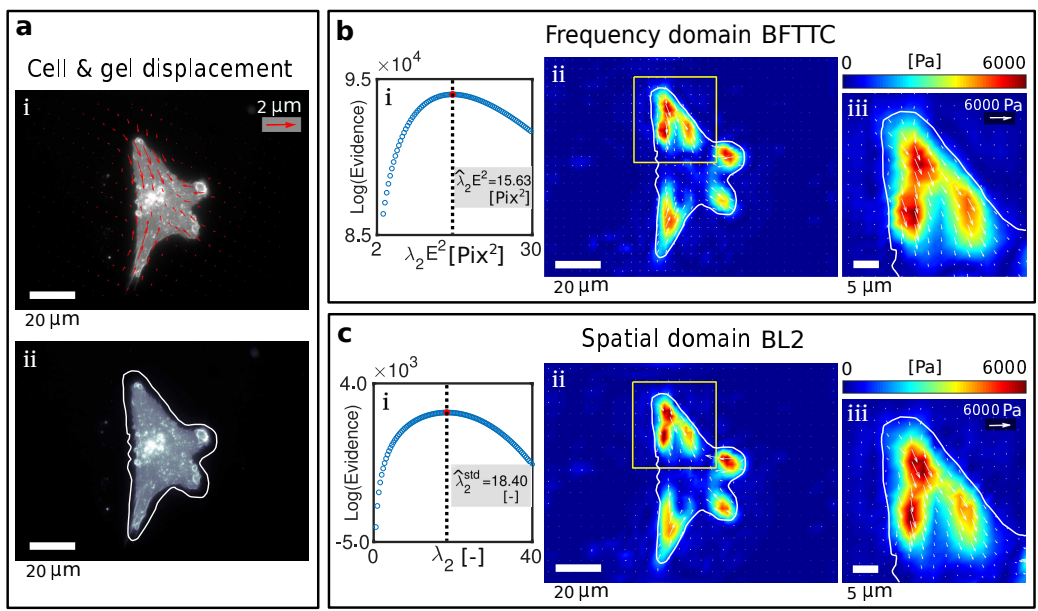}
		\caption{{\bf Test of Bayesian Fourier transform traction cytometry (BFTTC)  using experimental data.}
			(a,i-ii)~Image of an adherent cell and the measured gel displacements. Only every 7-th displacement is shown for better visibility. The cell edge is outlined in white. (b)~Results from traction calculation with BFTTC. (b,i)~A plot of the logarithmic evidence reveals a clear maximum, which serves to determine the regularization parameter. (b,i-ii)~Calculated traction forces.  (c)~Results from traction calculation with the real-space method BL2 for comparison with BFTTC. While the two methods produce similar results, traction fields calculated with BFTTC are slightly smoother than the fields calculated with the real-space method due to the different discretizations.}
		\label{fig5}
	\end{figure}

\begin{figure}[!h]
		\centering
		\includegraphics[width=1\linewidth]{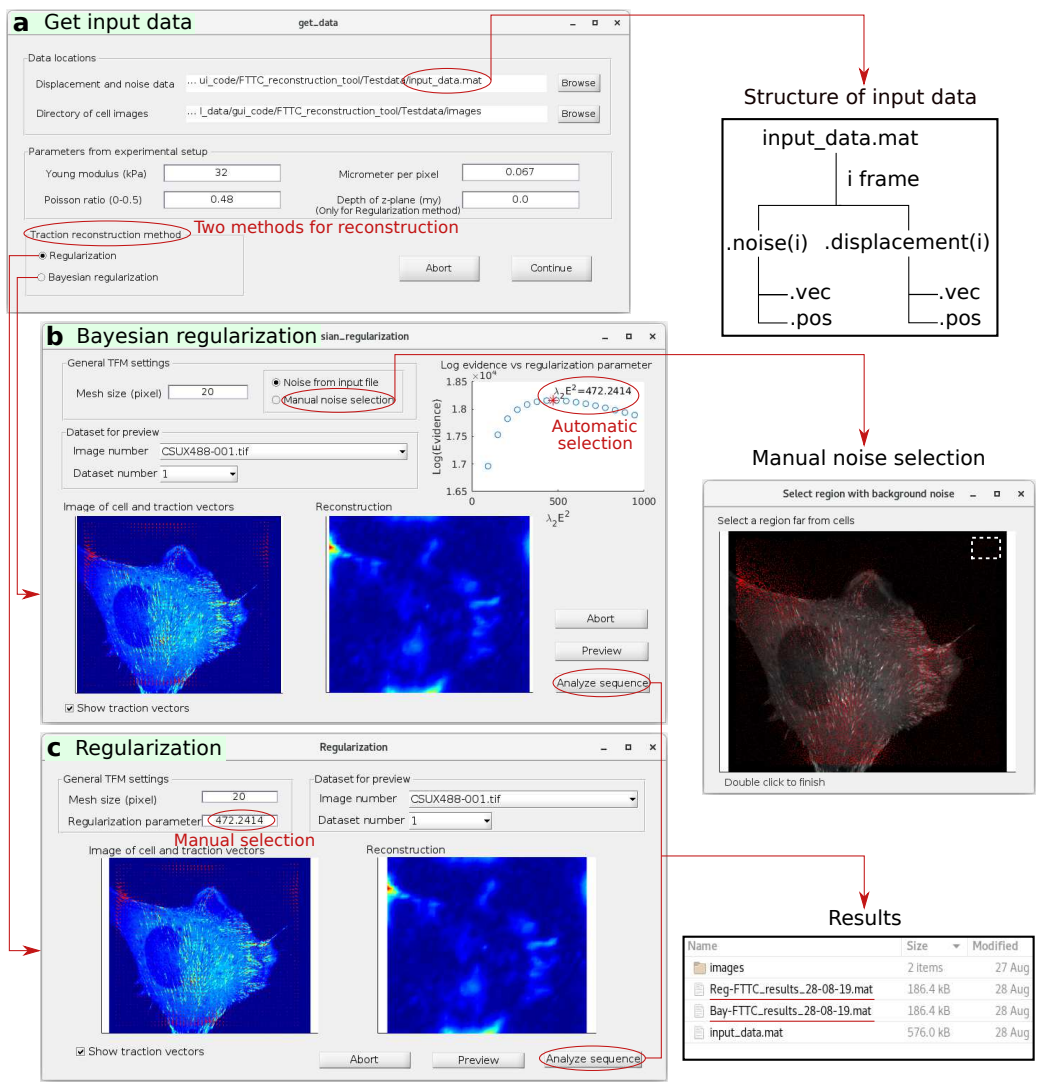}
		\caption{{\bf Graphical user interface of the provided software for regularized FTTC and BFTTC.}
			(a)~The ``get data'' interface allows users to input data locations and parameters of experimental setup. The data structure of the input files can handle a whole video sequence or individual traction recordings. (b)~If the ``Regularization'' option is chosen, a regularization parameter in units of $\mathrm{Pix}^2$ must be provided by the user. (c)~If the option ``Bayesian regularization'' is chosen, the regularization parameter is automatically determined from the measured displacement data and its noise variance. A sample with displacement noise can either be provided with the input file or it can be determined from a manually selected region that is far away from the cell. A ``Preview'' button offers the possibility to visually inspect the solution before one presses ``Analyze sequence'' to calculate and save the results.}
		\label{fig2}
	\end{figure}

To provide an application example for BFTTC, we quantified the traction forces generated by NIH 3T3 fibroblasts and mouse podocytes on polyacrylamide gel substrates. The
experiments were done precisely as described in Refs.~\cite{plotnikov2014high,Schell2018}. The gel substrates had a Young's modulus of $E=32\,{\rm kPa}$ and a Poisson's ratio of roughly
$\nu=0.48$. Figure~\ref{fig5}(a) shows a cell outline and the measured displacement data. After recording images of the
cell and the nanobeads, the cell was removed from the substrate to provide a stress-free reference for tracking the motion of the nanobeads. We estimate the variance of the noise in the displacement data by quantifying the displacement variance in a small region that is very far away from
the cell and contains no systematic displacement. Plotting the logarithmic evidence as a function of $\alpha$ yields a curve with a clearly defined maximum,
see Fig.~\ref{fig5}(b--i,c--i), which results in an unambiguous selection of the regularization parameter. Further tests were performed where,
for visualization of the force-generating structures, the cell--substrate adhesions were labeled with GFP-paxillin. It was found that BFTTC produces traction maps with defined foci that co-localize with the GFP-labeled sites of focal adhesion, see the example in Fig.~\ref{fig2}. 

	\section{Summary}
	Traction force microscopy is a popular technique for studying minute forces generated by biological cells, as well as
	wetting or frictional forces, on soft substrates. The technique is based on the measurement of substrate displacements below the specimen,
	which allows calculation of the traction forces. Usually, this calculation is done by solving an inverse linear problem involving elastic Green’s functions. The procedure 
	requires methods for noise suppression. Dealing with noise appropriately is an essential issue since the linear system can be ill-conditioned, which means that the noise can become amplified to an extent that the true solution is entirely degraded. A simple way to remove the effects of noise is to filter the displacement field prior to traction reconstruction. This strategy
	usually works if the linear problem is solved in Fourier space because the resulting linear system is sparse.
	An alternative strategy for dealing with noise is regularization, most popular is L2 regularization. 
	With L2 regularization, spatial high-frequency variations in the data are suppressed, which leads to a robust solution of the
	inverse problem of calculating the traction. Regularization is more versatile than data filtering since it can deal with higher levels of noise,
	works both in real-space and Fourier-space approaches, and ensures robust reconstruction if non-standard Green's functions are employed, 
	for example to take into account three-dimensional substrate topography and tractions. Regardless of the method, suppression of
	noise always reduces the spatial resolution. Optimal resolution of the fine details of the traction field can only be gained if 
	the level of noise suppression is adapted for each sample. For L2 regularization, this adaptation is done by changing the regularization parameter,
	which is usually a manual process based on heuristics, which introduces a considerable degree of subjectivity in the resulting traction.\\
	Here, we have introduced a Bayesian method for automatic inference of the L2 regularization parameter for traction reconstruction in Fourier space.
	Using synthetic data of different type, we demonstrate that Bayesian Fourier transform traction cytometry (BFTTC) is a fast and reliable method.
	Our tests show that BFTTC can handle large measurement noise. However, the noise- and displacement variances ideally satisfy $\sigma^2_{\text{noise}}/\sigma^2_{\mathbf{u}} \lesssim 0.01$ for accurate traction reconstruction. While the quality of traction reconstruction with BFTTC is comparable to other methods based on L2 regularization, the choice of the regularization parameter is now automated. Heuristics like the L-curve criterion, which is particularly ambiguous in Fourier space, are no longer required. The additional computation time required for determining the optimal regularization parameter in BFTTC is only a few seconds to minutes for large data sets. {In our experience, the logarithmic evidence always exhibits a maximum that is sufficiently pronounced to yield a regularization parameter estimate. However, it is important to keep in mind that the algorithm is based on the assumption of a Gaussian prior distribution that is symmetric around the origin. Thus, the use of BFTTC is not recommended if the traction forces in the field of view do not balance each other. Moreover, if complex, non-Gaussian traction distributions, e.g., multi-modal distributions, are expected, it may be preferable to resort to Bayesian methods with prior distributions tailored to the specific problem in order to maximize the reconstruction quality.}
\\ 
To provide users from biology, physics, and materials sciences with an easy-to-use tool to analyze their TFM data, we implemented BFTTC as well as regularized FTTC as a Matlab package. The package comes with a user-friendly graphical interface, requires minimal knowledge of the algorithmic details, and is freely available~\cite{downloadlinkforourprogram}. 
	
	\section{Acknowledgments}
	We thank S.~V. Plotnikov (University of Toronto) and C. Schell (Albert-Ludwigs-University Freiburg) for providing the experimental test data.

	
	
	%% References
	
	
	\bibliographystyle{elsarticle-num}
	
	

	\newpage
	
\end{document}